\documentclass[twocolumn,aps,superscriptaddress,floatfix,longbibliography]{revtex4-2}

\usepackage[dvipsnames]{xcolor}
\usepackage{amsmath}
\usepackage{amssymb}
\usepackage{graphicx}
\usepackage{comment}
\usepackage{bm}
\usepackage{braket}
\usepackage{ulem} 
\usepackage[caption=false]{subfig}
\usepackage{ragged2e}
\usepackage[colorlinks=true,bookmarks=false,citecolor=blue,linkcolor=red,hyperfootnotes=true,urlcolor=blue]{hyperref}
\usepackage{lipsum}

\newcommand{\be}{\begin{equation}}
\newcommand{\ee}{\end{equation}}
\newcommand{\bea}{\begin{eqnarray}}
\newcommand{\eea}{\end{eqnarray}}

\newcommand{\la}{\langle}
\newcommand{\ra}{\rangle}

\hypersetup{
 colorlinks=true,
 linkcolor=blue,
 anchorcolor = blue,
 citecolor = blue,
 filecolor = blue,
 urlcolor = blue
}

  \newcommand{\Eq}[1]{Eq.~\eqref{#1}}

  \newcommand{\Fig}[1]{Fig.~\ref{#1}}

  \newcommand{\rsym}[1]{\ensuremath{%
  \mathfrak{r}_{\mathrm{\ifx\@empty{}\else{#1}\fi}}}}

  \definecolor{cEdit}{rgb}{.17,.44,.76}

\begin{document}
	
\title{Detecting Multipartite Entanglement Patterns using Single Particle Green's Functions}
	
\author{Rajesh K. Malla}
\author{Andreas Weichselbaum}
\affiliation{Condensed Matter Physics and Materials Science Division, Brookhaven National Laboratory, Upton, New York 11973, USA}

\author{Tzu-Chieh Wei}
\affiliation{C.\,N. Yang Institute for Theoretical Physics and Department of Physics and Astronomy, State University of New York at Stony Brook, Stony Brook, NY 11794, USA}

\author{Robert M. Konik}
\affiliation{Condensed Matter Physics and Materials Science Division, Brookhaven National Laboratory, Upton, New York 11973, USA}
	
\begin{abstract}
{We present a protocol for detecting multipartite entanglement in itinerant many-body electronic systems using single particle Green's functions. To achieve this, we first establish a connection between the quantum Fisher information (QFI) and single particle Green's functions by constructing a set of witness operators built out of single electron creation and destruction operators in a doubled system. This set of witness operators is indexed by a momenta $k$.  We compute the QFI for these witness operators and show that for thermal ensembles it can be expressed as an auto-convolution of the single particle spectral function. We then apply our framework to a one-dimensional fermionic system to showcase its effectiveness in
detecting entanglement in itinerant electron models.  We observe that the detected entanglement level
is sensitive to the wave vector associated with witness operator.  Our protocol will permit detecting entanglement in many-body systems using
scanning tunneling microscopy and angle-resolved photoemission
spectroscopy, two spectroscopies that measure the single particle Green's function.  It offers the prospect of the experimental
detection of entanglement through spectroscopies beyond the established route of measuring the dynamical spin response.}


\end{abstract}
	
\date{\today}

\maketitle

{\it Introduction:} Multipartite entanglement is a fundamental feature of quantum physics that describes the entanglement of more than two subsystems, i.e., particles, or various degrees of freedom \cite{RMP-1, RMP-2}. It encodes a higher level of complex correlations as compared to bipartite entanglement and enables applications in quantum metrology \cite{MPE-precision2010,MPE-precision2012}, quantum communication and teleportation \cite{PRAcommunication-1, PRL2000-teleportation, Nature-teleportation}, as well as quantum computation \cite{Ekert1998,nc2010}. More generally in many-body physics,  quantum entanglement plays a central role in understanding emergent collective phenomena including quantum spin liquidity \cite{qsl-sci, QSL-report, QSL-RMP}, topological order \cite{Kitaev2006PRL,Xiao2006PRL,Balents2012topology,QSL-RMP}, and quantum criticality \cite{Criticality2007PRL,Criticality2007PRE,Criticality2007PRA,pro2020}. 

Detecting quantum entanglement beyond bi-partite entanglement \cite{tomography-1,tomography-2,interference} is challenging as the complexity and computational requirements increase with the number of subsystems.  The quantum Fisher information (QFI), initially proposed as a concept in quantum metrology~\cite{helstrom1969quantum,braunstein1994statistical,giovannetti2011advances,toth2014quantum}, has been employed to quantify multipartite entanglement \cite{PhysRevLett.102.100401,pezze2014quantum,hauke2016, QFIMPE1,PhysRevB.99.045117}. Exploiting experimentally accessible quantities that connect to the QFI has been one of the most promising strategies for detecting multipartite entanglement. In particular, dynamical susceptibilities have recently been proposed to be good candidates for quantifying multipartite entanglement \cite{hauke2016} and have been used in inelastic neutron scattering experiments on magnetic systems \cite{Neutron-PRR,Neutron-PRL,Neutron-PRB,Moore2023}.  While multipartite entanglement has been studied in itinerant electronic systems \cite{Hauke2021,PhysRevB.106.085110,Hales2023,X-ray-PRL}, the existing approaches do not connect single electron response functions to multipartite entanglement detection.  It is our purpose here to introduce new protocols that address this omission.

{In this Letter, we develop a theoretical protocol for
detecting {\it patterns} of multipartite entanglement via QFI in
itinerant electronic systems via single particle Green's
functions. This can enable experimental detection in platforms
where the spectral information is obtained from single electron
probes such as scanning tunneling microscopy (STM) and
angle-resolved photoemission spectroscopy (ARPES) experiments,
two techniques ubiquitous in the characterization of quantum
materials \cite{Zhang2022,RevModPhys.93.025006,Bian2021,STM}.
The QFI is always associated with a particular choice of witness operator.
When the most natural choice of witness operator connected to STM and APRES is made, in analogy to
the prior work on neutron scattering detection of entanglement \cite{Neutron-PRR,Neutron-PRL,Neutron-PRB,Moore2023}, the QFI
fails to detect any entanglement.
To circumvent this problem,
we design a witness operator via a non-trivial doubling
technique, where the witness operator acts across 
two copies of
the same original system that do not otherwise interact. As our
main result, we present a temperature-dependent expression for
QFI which directly depends on the single particle spectral
function. We also derive bounds for fine-grained entanglement
patterns corresponding to various combinations of subsystems and
demonstrate the role that symmetry plays in these bounds. Unlike
previous approaches \cite{hauke2016,Hauke2021}, where the bound
describes only the largest subsystem, our method considers
various subsystem combinations. The amount of detected
entanglement depends on the wave vector associated with the
witness operator. 
Comparing measured QFI values against QFI bounds over a
set of wave vectors, $0\leq k < 2\pi$, detects entanglement more
efficiently than focusing on any single $k$. We apply our
protocol to a one-dimensional (1-D) finite-sized itinerant
electron model, successfully excluding various entanglement
patterns at both zero and finite temperatures.  }

{\it Background:} The QFI of a state $|\psi\rangle$ relative to an observable $\mathcal{O}$, denoted by $F_Q(|\psi\rangle,\cal O)$,
is a fundamental concept in quantum metrology~\cite{helstrom1969quantum}. 
It puts a bound on the precision in
the quantum measurement of a phase $\theta$ of a unitary
operator, $U=e^{i\theta \hat{\cal O}}$, with $\hat{{\cal O}}$ being a Hermitian observable. 
The bound is known as
the quantum Cram{\'e}r–Rao bound,
$(\Delta\theta)^2\geq 1/[mF_Q(\hat{\cal O},\rho)]$, where $m$ is
the number of independent measurements used to infer $\theta$.  
For a mixed state 
$\rho = \sum_{\lambda} p_{\lambda} |\lambda\ra \la\lambda|$
the QFI is defined by 
$F_Q(\rho,\hat{\cal O}) = 2\sum_{\lambda \lambda'}
    \frac{(p_{\lambda}-p_{\lambda'})^2}{p_{\lambda}+p_{\lambda'}}
    |\langle \lambda |
    {\hat{\cal O}} |\lambda'\rangle|^2$.
For a pure state $|\psi\rangle$, this simplifies to the
variance $F_Q(\rho,\hat{\cal O}) = 4(\langle\psi|\hat{\cal
O}^2|\psi\rangle-\langle \psi |\hat{\cal O}|\psi\rangle^2)\equiv
4 \Delta^2 {\cal O}$. 
From this, it is apparent  that the QFI is tightly related to 
fluctuations, and thus
the precision with which a quantum state $\rho$ can be used to estimate the angle $\theta$.

The value of the QFI for a given state is related to the
amount of multipartite entanglement it possesses
\cite{PhysRevLett.102.100401}, but
is nonetheless highly sensitive
to the choice of witness operator ${\cal O}$ \cite{hauke2016,
QFIMPE1,pezze2014quantum}. The general form of the witness
operator that we consider here is ${\hat {\cal O}}=\sum_{i}
{\hat {\cal O}}_i$, where ${\hat {\cal O}}_i$ denotes a local
operator acting on-site $i$ and whose spectra lies within the
interval $(h_{i,\rm min},h_{i,\rm max})$.  
{\it If no symmetries are taken into account}, $(h_{i,\rm min},h_{i,\rm max})$ determine the bounds for which the QFI of a quantum state $\rho$ must exceed in order for a certain amount of entanglement to be detected.  Specifically, $\rho$ is at least ($m+1$)-partite entangled (with $m$ being an integer) if its QFI satisfies \cite{Hauke2021}
\begin{eqnarray}\label{simple_bound}
F_Q(\rho,{\hat {\cal O}}) &>& m {\sum_{i=1}^N} 
   (h_{i,\rm max}-h_{i,\rm min})^2,
\end{eqnarray}
{with  
$N$ the number of sites.  While \Eq{simple_bound} is appealing in its simplicity, we will see that it cannot be directly used.}

{{\it Construction of witness operators:}}
{As a first step,} we turn to the problem of building a
witness operator, ${\cal O}$, out of single fermion operators.
While this might appear straightforward, the naive choice, as just advertised,
${\cal O}_i=c^\dagger_i + c_i$, does not work. Here with
$h_{i,{\rm max}/{\rm min}}=\pm 1$ and ${\cal O}_i{\cal O}_j+{\cal
O}_j{\cal O}_i=2\delta_{ij}$, $F_Q({\cal O})
= 4 (N - \langle \psi |\hat{\cal O}|\psi\rangle^2)
\leq 4N$.
According to the simple bound in \Eq{simple_bound} this fails
to detect any entanglement.  One strategy might then to be to take into account symmetries in the bound, such as overall electron number, thus invalidating \Eq{simple_bound}.
However, if we do so we find that $F_Q({|\psi\rangle,\cal O})$ is {\it identically} $4N$
regardless of quantum state $|\psi\rangle$ and so again cannot
discriminate between levels of entanglement.  In part, the
failure of the naive choice arises because the witness operator
is fermionic and so has a trivial anti-commutation relation. 

To get around this limitation while still maintaining a
connection to the single particle response function, we create
two identical copies of our system, denoted as $A$ and $B$. The
copies do not interact with one another.  The density matrix of
the doubled system, denoted as $\rho_D$, is thus defined as the
tensor product of the original density matrix, $\rho$, for each
copy, yielding $\rho_D = \rho_A \otimes \rho_B = \rho \otimes
\rho$. Because the two copies are non-interacting, the
Hamiltonian of the doubled system, denoted as $H_D$, is
expressed as $H_D = H_A \otimes I_B \oplus I_A \otimes H_B$,
where $H_A = H_B = H$ represents the Hamiltonian of the original
system, and $I_{A, B}$ are the identity matrices corresponding
to the Hilbert space of the respective copies. Most importantly,
the dynamical information of the original system completely
determines the dynamics of the doubled system, as the two copies
are independent of one another.

Within this framework, we propose that operators that hop between the two copies can serve as promising candidates for detecting multipartite entanglement.  Such operators, involving fermion bilinears, are bosonic.  Specifically, 
for an $N$-site spinless electron system
the witness operator can be expressed as
\begin{eqnarray}
\hat{\cal O}(\bar{a}) &=& \sum_{j=1}^{N} a_{j} \hat{c}_{j A}^\dagger \hat{c}_{j B} + \text{H.c.}
\label{WO_SPGF}
\end{eqnarray}
{with $\hat{c}_{j\mu}^{(\dagger)}$ 
the annihilation (creation) operator at site $j$
for copy $\mu \in \{A,B\}$. 
We assume local hopping across copies $A$ and $B$, with
$a_i \in \mathbb{C}$ the hopping strength w.r.t. site $i$,
denoting $\bar{a} = (a_1,\ldots, a_N)$.}
This {diagonal structure} ensures our QFI satisfies additivity
{e.g., for disconnected system blocks}.

To turn $F_Q({\cal O})$ into an effective detector of
entanglement, we need to do more beyond simply doubling the
system.  For instance, if we use this witness operator without
imposing any symmetry constraints, we see that in order to
detect ($m+1$)-partite entanglement, the simple bound based on
the local min/max eigenvalues requires $F_Q$ to exceed $4Nm$ as
the min/max eigenvalues of the hopping operator are $\mp 1$.
However,
{for normalized $\bar{a}^2 = N$
such as $|a_i|=1$,  
\Eq{WO_SPGF} results in}
$F_Q({\cal O}) {\leq 4N}$.
{This strict upper bound is a consequence of the
fermionic nature of the witness operator \eqref{WO_SPGF}
w.r.t. a particular system $A$ or $B$ \cite{SM}.
This prevents
multipartite entanglement detection
via \Eq{simple_bound} in a conventional sense}.

{However, if we now take symmetries into account, $F_Q({\cal O})$ can be turned into a useful detector of entanglement.  We have examined 
how symmetries alter}
the entanglement
bounds for three different symmetries:
i) overall electron number; ii) electron number parity; and iii)
time reversal invariance.  We treat case i) in the main text of
this letter, leaving case ii) and iii) to S4 of \cite{SM}.  We
will see that electron number as a good quantum number can
reduce the bound that the QFI has to exceed to see a certain
amount of entanglement.  

To see how the symmetries affect the entanglement bounds that the QFI has to exceed in order to detect a certain level of entanglement, let us consider a toy example of a two-site system with spinless fermions. {Assuming that}
the electron number is a good quantum number, {the
one-fermion states can be written as $|\psi(u_1,u_2)\rangle =
(u_1 c_1^\dagger + u_2 c_2^\dagger) |0\rangle$, 
{with coefficients $u_1, u_2 \in \mathbb{C}$ 
and $|0\rangle$ the vacuum state.}
When one of the two coefficients is zero, then the state has 1-partite entanglement, i.e., is unentangled. When both of the coefficients are non-zero the wavefunction has 2-partite entanglement \cite{Hauke2021}.}
To detect this entanglement, we examine the QFI corresponding to our witness operator $\hat{{\cal O}}$ in 
\Eq{WO_SPGF}. For any choice of $a_1$ and $a_2$, the QFI for the states $|\psi(u1,0)\rangle$ and $|\psi(0,u2)\rangle$ is zero.  However, for $|\psi(u1,u2)\rangle$, it equals $F_Q =
8 |a_1 u_1|^2 + 8 |a_2 u_2|^2 - 8 |a_1u_1^2 + a_2u_2^2|^2$. 
Notably, when $u_1 = u_2 = 1/\sqrt{2}$ (nominally the maximally entangled state in the one-electron sector), the QFI simplifies to $F_Q(|\psi(1/\sqrt{2},1/\sqrt{2})\rangle,\hat{{\cal O}}) =
{2|a_1-a_2|^2}$. 
This analysis reveals that a suitable witness operator for a two-site system requires $a_1 \neq a_2$. Provided this is so, our witness operator successfully distinguishes between one-partite entanglement ($F_Q = 0$) and two-partite entanglement ($F_Q \neq 0$) in this toy two-site model. Setting $a_1 = 1$ and $a_2 = 0$ yields a one-site witness operator, but one still capable of detecting entanglement between sites, showing that spatial entanglement can be detected, in principle, by local probes like STM. 

{{\it Main result:} 
As the central result of this work, we derive an exact expression for the QFI in terms of a single particle spectral function for the witness operator in Eq.~\ref{WO_SPGF} with $a_j(k)=e^{ikj}$,
\begin{align}
   & \hspace{5ex} F_Q(k,T) = \frac{2}{\pi} \int^\infty_{-\infty} 
   d\omega \tanh(\tfrac{\beta\omega}{2}) \, S(\omega,k,T)
\label{finT_QFI}\\ 
   & \text{with } S(\omega,k,T) = {\rm Im}\int^\infty_0 e^{i\omega t}\,
     i \langle [\hat{\cal O}(k,t),\hat{\cal O}(k,0)]\rangle_T
\notag\\
   &
     \!=\! \tfrac{2e^{\beta\omega}}{n_B(\omega)} 
   \sum_q \!\int \! d\omega'\,
      n_F(\omega')
      n_F(-\omega{-}\omega')
      A_q(\omega')
      A_{q+k}(\omega{+}\omega'), \quad
\notag
\end{align}
where $T$ is the temperature, $S(\omega,k,T)$ is the imaginary
part of the dynamical susceptibility of the density matrix for
the witness operator ${\cal O}(k)$ \cite{hauke2016}, and
$n_{B/F}(\omega)=(e^{\beta\omega}\mp 1)^{-1}$ are the
bosonic/fermionic filling functions respectively.  Here the
relevant susceptibility, $S(\omega,k,T)$, is a weighted
convolution of two copies of the single particle spectral
function, $A_q(\omega)$, at wavevector $q$ and energy $\omega$.
$A_q(\omega)$ is a quantity readily accessed in STM and ARPES
experiments. The above result allows one to take any measurement
of $A_q(\omega)$ and convert it into a bound on the amount of
multipartite entanglement in a material. See S1A of \cite{SM}
for an explicit derivation.  

The expression for the QFI  for a pure state $|\psi\rangle$ with
a definite electron number (DEN) of $N_e$ electrons reads
\begin{equation}
F_Q(|\psi\rangle,\hat{{\cal O}}(k))
  =  8N_e- 8\sum_{i,j=1}^{N} \cos(k(i-j))
   \,|C_{ij}|^2
\label{ps_QFI}
\end{equation}
{with $C_{ij} \equiv \langle \psi| c_{i}^{\dagger}c_{j}|\psi\rangle$
the 1-particle correlation matrix.} 
At zero temperature, \Eq{finT_QFI} reduces to \Eq{ps_QFI}.
We consider more general forms of $F_Q$ where the wavefunction,
$|\psi\rangle$, has an indefinite electron number (IEN) in S2 of
\cite{SM}.  This QFI is {\it additive} as ${\cal O}(k)$ involves
hopping between the same sites $i$ of copies A and B: if a
wavefunction can be represented as the tensor product of $M$
blocks, $
    |\psi(N)\rangle = |\psi(N_1)\rangle \otimes |\psi(N_2)\rangle \otimes \cdots \otimes |\psi(N_M)\rangle ,
$
with $\sum^M_{i=1} N_i = N$, 
the QFI for $|\psi(N)\rangle$ satisfies
$
    F_Q(|\psi(N)\rangle,{\cal O}(\bar a(k)) = \sum_{i=1}^M F_Q(|\psi(N_i)\rangle,{\cal O}(\bar a(k))) .
$
While the QFI for this witness operator has $4N$ as a strict upper bound, we will now show that for wavefunctions with DEN, the bounds for particular entanglement {\it patterns} (to be explained) as a function of the wavevector $k$ are less restrictive.  Here it will be important to consider $F_Q(k,T)$ at all $k$, not a single $k$, to effectively detect such patterns.

}

\begin{figure}
\centering
\includegraphics[scale=0.19]{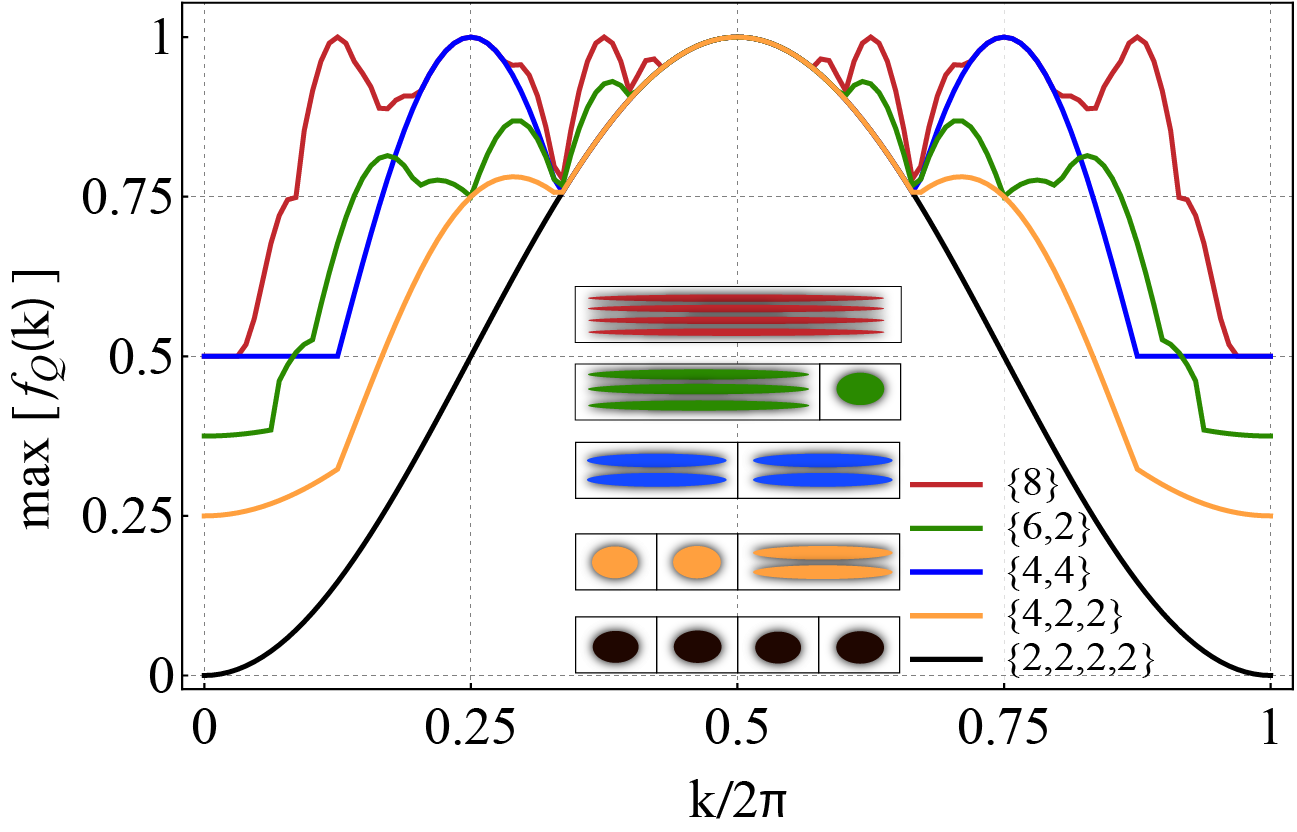}
\caption{The maximal QFI for various entanglement patterns is plotted as a function of $k$ for a $8$-site spinless fermionic system at half-filling, i.e., $N_e=4$. The entanglement patterns are shown schematically in the inset. The system is split up into independent blocks where electrons are confined within a block as visualized by the colored clouds. {The smallest block size here contains 2 sites and the bars separating individual sites have been omitted within a block.}}
\label{fig:1}
\end{figure}

{{\it Fine-grained entanglement patterns:}}
We first need to find the maximal attainable value of $F_Q(|\psi\rangle,{\cal O}(k))$ at each value of $k$. Although a full analytical determination of this is unavailable, we can numerically compute it for finite $N$. Similar to the 2-site case, a state with 1-partite entanglement has an $F_Q$ that is identically zero.  Therefore, any finite value of $F_Q$ indicates that the wavefunction has at least two-partite entanglement. We now turn to determine the maxima for states with at least 2-partite entanglement. 

We start with entanglement patterns in an 8-site 
system at {half-filling} with DEN.  Five such
patterns for half-filling are given in \Fig{fig:1}: $\{8\}$, $\{6,2\}$, $\{4,4\}$, $\{4,2,2\}$,
$\{2,2,2,2\}$.  Here the notation $\{z_1,z_2,...\}$ indicates
that we have a wavefunction that decomposes into a block with
$z_1$-partite entanglement plus a block with $z_2$-partite
entanglement, and so on.  We exclude here patterns involving
1-partite entangled subblock
: $\{ 6,1,1\}$, $\{4,2,1,1\}$,
$\{4,1,1,1,1\}$, $\{2,2,2,1,1\}$, $\{2,2,1,1,1,1\}$, and
$\{2,1,1,1,1,1,1\}$, 
{and subblocks with an odd number of sites.}  Subblocks with 1-partite entanglement with DEN contribute 0 to the QFI and have maxima equal to those for smaller systems, {while the subblocks with an odd number of sites will not satisfy a half-filling condition.} 
We have determined numerically what the maximal QFI density, $f_Q(k)$,
defined as
$f_Q(k)\equiv F_Q({\cal O}(\bar a(k)))_{\text{max}}/4N$,
is for each entanglement pattern and
reported it in \Fig{fig:1}.

 \begin{figure}
\centering
\includegraphics[scale=0.18]{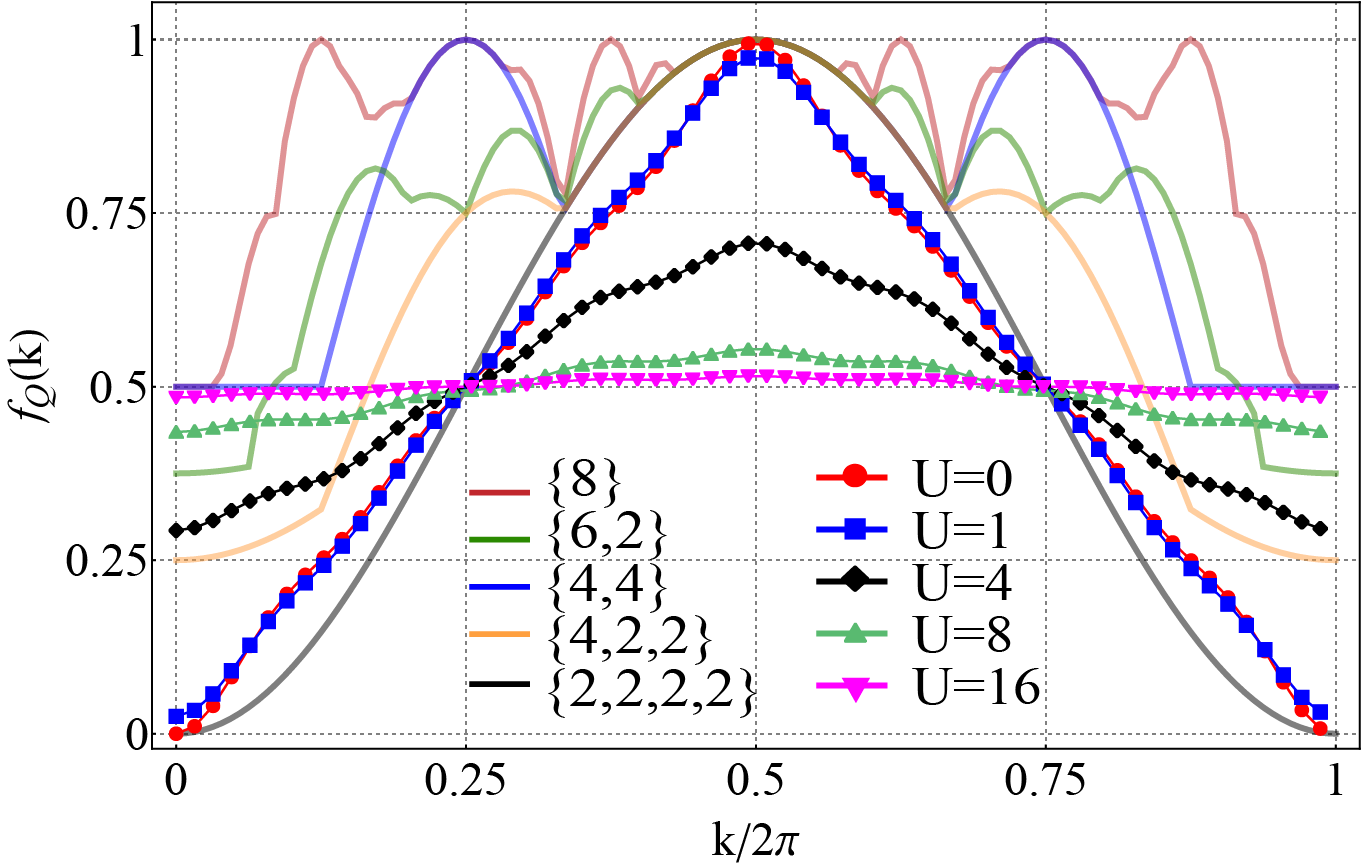}
\caption{The QFI of ground state wavefunctions for various interaction strengths 
are shown for the half-filled $t-U$ Hamiltonian (\ref{t-U}) with $N=8$ sites (plot markers), alongside the maximal QFI curves for various entanglement patterns 
(solid lines; same as in \Fig{fig:1} for reference). The wavefunctions are obtained for periodic boundary conditions (PBC) and $t=1$. 
}\label{fig:2}
\end{figure}


In Fig.~\ref{fig:1} we see the maximal QFI density for each
pattern is a non-trivial function of $k$ that possesses
reflection symmetry about $k=\pi$.  As already stated, the maximal possible value of the QFI is $4N$ at any k.  This maximum is obtained at the commensurate wavevectors $k=2\pi n/N$, where $n=1,\cdots,N-1$.  At $k=0,2\pi$ the maximal QFI value is $2N$, 1/2 the maximal possible value.  We provide insights as to why these peaks occur at these wave vectors in S3 of~\cite{SM}. These
maximal values do not depend on the ordering of the subblocks
within a pattern.  For example, a pattern $\{4,2,2\}$ has the
same maxima as the patterns $\{2,4,2\}$ or $\{2,2,4\}$. 

We also see from Fig.~\ref{fig:1} the effects of having a 2-partite subblock on what the maximal QFI is. 
 At $k=0,2\pi$, such 2-partite subblocks give 0 contribution to the maximal QFI. However, if a subblock has size
$n>2$, 
its contribution to the maximal QFI at $k=0,2\pi$ is $2n$. Correspondingly, if a pattern has $n$ 2-partite subblocks its maximal value at $k=0,2\pi$ is $4(N/2-n)$.  
 
We stress here that in this way of computing the maximal QFI, we
are not merely providing the data to determine if there is an
entangled subblock of a certain size in the system.   If the
bound for a particular pattern in \Fig{fig:1} is exceeded for a
particular wavefunction, the entire pattern is excluded.  So
while $\{4,4\}$ and $\{4,2,2\}$ both have 4-partite entangled
 blocks, we can nonetheless, using the bounds of \Fig{fig:1},
 distinguish between the two.  {The distinguishability between these fine-grained entanglement patterns is a key result.}

Fig.~\ref{fig:1} is constructed for an $N=8$ site system.  However, it has
greater applicability to much larger system sizes.  If we have a
system with $8K$ sites, $K>1$, and its block structure is a
repetition of one of the patterns in \Fig{fig:1}, i.e.,
$\{8,\cdots,8\}$ with $\{8\}$ repeated $K$-times, then the maximal QFI density
for this pattern in the $8K$-site system is identical to the $8$-site pattern shown in  \Fig{fig:1}.  The replication of a particular 8-site
pattern only affects the maximal QFI through a trivial scale
factor.  This result follows from the additivity of the QFI -- see S5 of \cite{SM}.

 
 \begin{figure}
\centering
\includegraphics[scale=0.08]{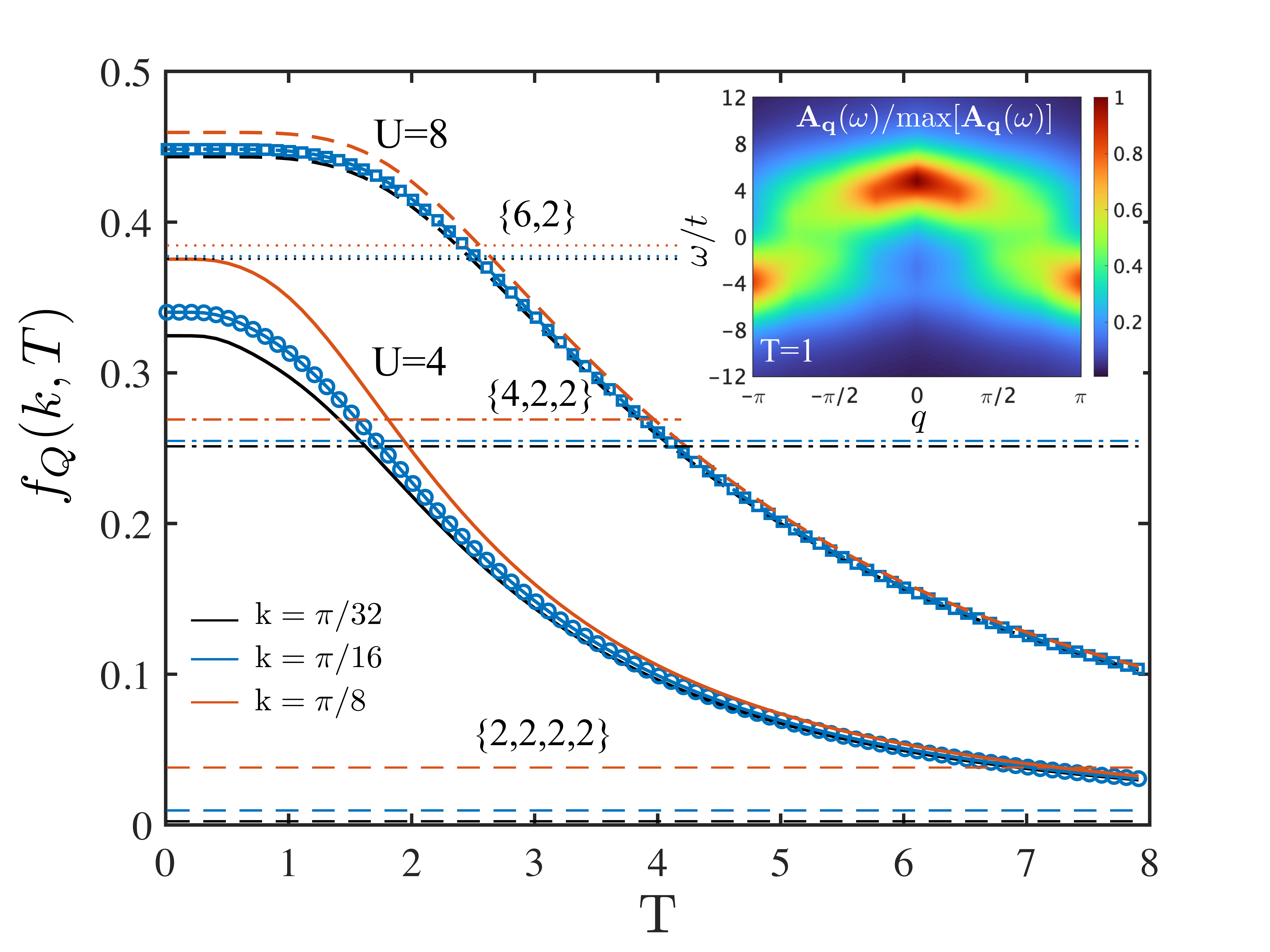}
\caption{ Temperature dependence of the QFI for the $N=8$ $t - U$ model in Eq. (\ref{t-U}) for two values of interaction
    strength, $U=4$ (lower set of curves) and $U=8$ (upper set
    of curves).
    The horizontal lines mark the maximal possible QFI for various
    entanglement patterns. 
    The three curves for each $U$ or entanglement
    pattern correspond to different momenta as specified by color in the legend. For visual clarity, we have added markers to only one curve.
    The spectral function $A_{q}(\omega)$ normalized by its maximum
    is shown in the inset for $U=4$ and $T=1$.
    The spectral data in the inset 
    is broadened in energy with a Lorentzian 
    of width $\eta=2t$ - see S1B of \cite{SM}.
}\label{fig:3}
\end{figure}

{{\it Application to 1-D fermionic system:}} Having established the maximal QFIs for a variety of different entanglement patterns and how to compute the QFI of a model both at zero (Eq. \ref{ps_QFI}) and finite (Eq. \ref{finT_QFI}) temperatures, we now turn to applying this formalism to a particular itinerant electron model, the $t-U$ Hamiltonian for spinless fermions,
\begin{equation}
    H = t\sum_{i}(c^\dagger_{i} c_{i+1}+c^\dagger_{i+1}c_{i}) + U\sum_i \hat n _{i} \hat n_{i+1}.
    \label{t-U}
\end{equation}
Here $t$ is the nearest neighbor hopping strength and $U$ is the nearest-neighbor Coulomb interaction.

In \Fig{fig:2}, we plot the normalized QFI, $f_Q(k)$, as a function of wavevector, $k$, for different ground state
wavefunctions corresponding to different values of $U$ alongside the QFI maxima
obtained in \Fig{fig:1}. The QFIs for both ground state wavefunctions and excited state wavefunctions for any interaction strength, including $U=0$, lie above the maximal QFI for \{2,2,2,2\} entanglement pattern. The higher entanglement pattern \{4,2,2\} is excluded for the ground state wave functions with interaction strengths $U=4$, $8$, and $16$, while the \{6,2\} pattern is excluded when the interaction strengths are $U=8$ and $16$. The results in \Fig{fig:2} show that the pattern \{4,4\} is not excluded for any ground state wavefunctions.

In \Fig{fig:3} we turn to finite temperature and plot the QFI of the $t-U$ Hamiltonian as
a function of temperature for several different discrete momenta $k$ and interaction strengths $U$. Here the QFI is computed by first computing $A_{q}(\omega)$ (the inset of \Fig{fig:3} shows this quantity at $U=4, T=1$) and then employing Eq. \ref{finT_QFI}.
The QFI curves in \Fig{fig:3} are compared with the
reference lines that mark the maximal QFI for various
entanglement patterns at particular $k$ values. The QFI
decreases with increasing temperatures and at some critical temperatures will begin to fail to
detect that pattern.    
For
instance, in the case of $U=8$, at temperature $T\sim 3$, the QFI of all three wavevectors considered
falls below the maximal QFI of the $\{6,2\}$ pattern, and at temperature $T\sim 4$, the QFI falls below the maximal QFI of the $\{4,2,2\}$ pattern.
At high temperatures, the
QFI falls below the $\{2,2,2,2\}$ for some of the $k$ values in the case of $U=4$. To exclude the $\{2,2,2,2\}$ pattern there, an effective strategy is to compute the QFI at smaller $k$ values as the exclusion bound for this pattern approaches zero as $k\rightarrow 0$.

In conclusion, we have developed a protocol by which information
contained in measurements of the single particle response
function can be converted into entanglement bounds.  This
entails the analysis of a fictitious doubled system together
with the determination of entanglement bounds that are not based
on single site eigenvalues of a single local witness operator,
but that are tied to a continuum set of witness operators
(parameterized by a wavevector $k$) and multipartite
entanglement patterns.  Rather than bounds determining whether a
single block of sites is entangled, we find bounds that exclude
a particular partition of entanglement across the entire system.
While we have focused on spatial entanglement, our method is
agnostic as to the degrees of freedom being entangled.  We can
easily with our approach, for example, work in $k$-space, and
ask whether entanglement exists across $k$-modes.

{\it Acknowledgments}.-- 
All authors of this work were supported in its production by the U.S. Department of Energy, Office of Basic Energy Sciences, under Contract No. DE-SC0012704.  RMK performed part of the work for this paper while visiting the Aspen Center for Physics, a center supported by the National Science Foundation under grant PHY-2210452.

\bibliographystyle{apsrev4-2}
\bibliography{ref,bib}

\end{document}


\title{Supplemental material for ``Detecting Multipartite Entanglement Patterns using Single Particle Green's Functions"}

\author{Rajesh K. Malla}
\author{Andreas Weichselbaum}
\affiliation{Condensed Matter Physics and Materials Science Division, Brookhaven National Laboratory, Upton, New York 11973, USA}

\author{Tzu-Chieh Wei}
\affiliation{C.\,N. Yang Institute for Theoretical Physics and Department of Physics and Astronomy, State University of New York at Stony Brook, Stony Brook, NY 11794, USA}

\author{Robert M. Konik}
\affiliation{Condensed Matter Physics and Materials Science Division, Brookhaven National Laboratory, Upton, New York 11973, USA}

\begin{abstract}
    Here we report some additional technical details on our work. In Section S1 we provide the explicit expression for the finite temperature QFI in terms of the single particle spectral function.  In Section S2 we provide the zero temperature expression for the QFI in terms of a generalized correlation matrix.  We then in Section S3 provide some insight using the correlation matrix as to why peaks in the maximal QFI curves occur at certain wavevector values of $k$. In Section S4 we consider the role of symmetries on QFI bounds.  Finally in Section S5, we extend our analysis for maximal QFI to larger systems.
\end{abstract}

\date{\today}
\maketitle

\renewcommand{\thesection}{S\arabic{section}}


\section{The QFI at Finite Temperature}

\subsection{Expressing $F_Q$ in terms of the single particle spectral function at finite temperature \label{sec:1}}

At finite temperature, the quantum Fisher information (QFI) can be written in terms of a weighted integral of the imaginary part of the retarded response function of the witness operator ${\cal O}(k)$ \cite{hauke2016}:
\begin{eqnarray}    
F_Q(\hat{{\cal O}(k)})
&=& \frac{2}{\pi} \int\limits^\infty_{-\infty} d\omega \tanh(\frac{\omega}{2T})S(\omega,k,T);\cr\cr
S(\omega,k,T) &\equiv& {\rm Im} \int\limits^\infty_0 e^{i\omega t}\, i\langle
   [ \hat{{\cal O}}(k,t), \hat{{\cal O}}(k,0)]\rangle_T .
\label{dyn-susc}
\end{eqnarray}
$S(\omega,k,T)$ is the
spectral density of the operator ${\cal O}(k)$ and can be written
in terms of 
the eigenstates $H|m\rangle=E_m|m\rangle$ of the Hamiltonian $H$ of the system,
\begin{eqnarray}
    S(\omega,k,T) &=& \pi(1-e^{-\beta \omega})\sum_{mn}
    \tfrac{e^{-\beta E_m}}{Z} |{\cal O}_{mn}|^2 
    \,\delta(\omega-E_n+E_m).\label{spec_den}
\end{eqnarray}
as well as the matrix elements ${\cal O}(k)_{mn}\equiv\langle m|{\cal O}(k)|n\rangle$ of the witness operator.  Here $Z=\sum_m e^{-\beta E_m}$ is the partition function.
We have defined the retarded response function so that the imaginary part is positive for $\omega>0$.

The explicit form of the witness operator
that we consider 
in the main text [Eq.\,(2)] 
has the general form
\begin{equation}
   \hat{\cal O}(\bar{a}) = \sum_{j=1}^{N} a_{j} \hat{c}_{j A}^\dagger \hat{c}_{j B}^{\,}
   + {\rm H.c.},
\label{WO_SPGF}
\end{equation}
where $\bar{a} = (a_1,\cdots, a_N)$ is a vector describing the hopping strength at each site, with $\hat{c}_{j A/B}^\dagger$ and $\hat{c}_{j A/B}$ the creation and annihilation operators at site $j$ for copies $A$ and $B$. Plugging in Eq. (\ref{WO_SPGF}) in Eq. (\ref{spec_den}) and removing indices $A$ and $B$ (as the copies are identical), we obtain the spectral function $S(\omega)$ of the doubled system as:
\begin{eqnarray}
S(\omega,T) &=& \frac{1-e^{-\beta\omega}}{4\pi}\sum_{ij}\int d\omega' \big[a_i a_j^* G^<_{ji}(-\omega')G^>_{ij}(\omega-\omega')
 +a_i^* a_j G^>_{ij}(\omega')G^<_{ji}(\omega'-\omega)]\big.
    \label{img-xi1}
\end{eqnarray}
Here $G^>/G^<$ are the greater/lesser single-particle Green's functions and are defined as 
\begin{eqnarray}
  G^>_{ij}(\omega) &=& 
  -i\int\limits^\infty_{-\infty} dt e^{i\omega t}\langle c_{i}(t)c^\dagger_{j}(0)\rangle 
 = \tfrac{-2\pi i}{1+e^{-\beta\omega}} 
   \,A_{ij}(\omega);
\\
   G^<_{ij}(\omega) &=&
   i\int\limits^\infty_{-\infty} dt e^{i\omega t}\langle c^\dagger_{j}(0)c_{i}(t)\rangle 
 = \tfrac{2\pi i}{1+e^{\beta\omega}} \,
   A_{ij}(\omega),
\label{Green'sfunctions}
\end{eqnarray}
where $A_{ij}(\omega)$ 
is the single particle spectral density
\begin{eqnarray}
    A_{ij}(\omega,T) &=& (1+e^{-\beta\omega})
    \sum_{nm} \tfrac{e^{-\beta E_m}}{Z}
    \langle m|c_i|n\rangle \langle n|c^\dagger_j|m\rangle
    \delta(\omega-E_n+E_m),
    \label{eq:spectral}
\end{eqnarray}
where here the eigenstates $|n\rangle$ are those of a single copy of the system.
It satisfies the sum rule 
$\int d\omega A_{ij}(\omega) = \{c_i,c^\dagger_j\} = \delta_{ij},
$, 
i.e., its integrated form reduces to the fermionic anti-commutator.  
The quantity $A_{ij}(\omega)$ can be measured in both angle-resolved photoemission and scanning tunneling microscopy experiments. 

If we now specialize the above formula to the case we are focused on in this paper, $a_j(k)=e^{ikj}$, $S(\omega,k,T)$ can be written as 
\begin{equation}\label{Somega}
    S(\omega,k,T) = 2\pi(1-e^{-\beta\omega})\sum_k \int^\infty_{-\infty}d\omega' n(\omega')n(-\omega-\omega')A_k(\omega')A_{q+k}(\omega+\omega'),
\end{equation}
where $n(\omega)=(1+e^{\beta\omega})$ is the Fermi filling function at energy $\omega$ and temperature $T=\beta^{-1}$.
Here we have introduced the Fourier transform of the single particle spectral densities:
\begin{eqnarray}\label{Aq}
    A_{ij}(\omega,T) &=& \frac{1}{N}\sum_q e^{{\rm i}q(i-j)}A_q(\omega,T);\cr\cr
    A_{q}(\omega,T) &=& (1+e^{-\beta\omega})
    \sum_{nm} \tfrac{e^{-\beta E_m}}{Z} \langle m|c_q|n\rangle \langle n|c^\dagger_q|m\rangle\delta(\omega-E_n+E_m), ~~~ c^\dagger_j = \frac{1}{N}\sum_q c^\dagger_q e^{{\rm i}qj}.
\end{eqnarray}
In writing the above, we are assuming the system is translationally invariant and momentum is a good quantum number.

\subsection{Additional plots for the QFI and spectral functions at finite temperature}
The temperature-dependent QFI for the thermal state of the $t - U$ Hamiltonian plotted in the main text is computed using
Eq. (\ref{dyn-susc}) and Eq. (\ref{Somega}). For finite system sizes where $A_q(\omega)$ is formally a sum over $\delta$-functions, we divide the range of $\omega$ over which $A_q(\omega)$ is non-zero into bins $(\omega_i,\omega_{i+1})$. A $\delta$-function, $\delta(\omega-b)$, whose weight falls in the interval $(\omega_i,\omega_{i+1})$ is replaced in Eq. \ref{Aq} by a product of Heaviside step functions: $\Theta(b-\omega_i)\Theta(\omega_{i+1}-b)/(\omega_{i+1}-\omega_i)$.  The range over which $A_{q}(\omega)$ appearing in Eq. (\ref{Somega}) is potentially nonzero is $[-\omega_0,\omega_0]$, where
$\omega_0=E_{max}-E_{0}$ represents the full many-body
energy bandwidth, 
with $E_{max}$ 
the energy 
of the highest excited eigenstate and $E_0$ 
the ground state energy.
This interval is then uniformly binned with a constant 
bin size of $\Delta \omega=0.1t$. The QFI is computed using the
binned data without broadening. 

\begin{figure}
    \centering
    \includegraphics[scale=0.2]{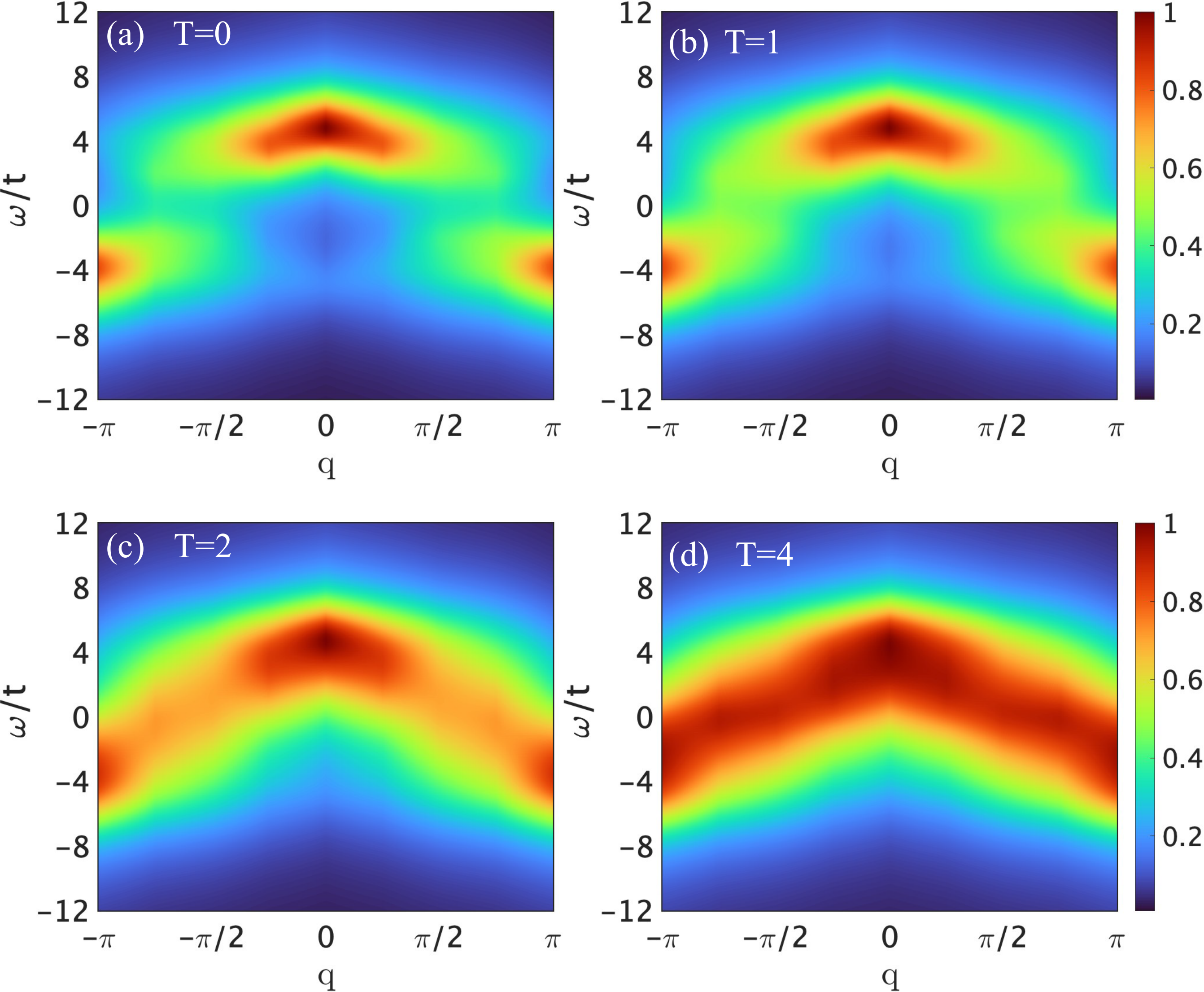}
    \caption{Shown are the spectral functions $A_{q}(\omega)$ for the $t-U$ Hamiltonian for $U=4$, which is computed from Eq. \ref{eq:spectral} and is normalized by its maximum. The broadening along $\omega$ is achieved by replacing the $\delta$-functions in the expression for $A_q(\omega)$ with broadened Lorentzians while the broadening along the $q$ axis is achieved using the interpolation tool in Matlab.
  Here 
    $q$ corresponds to the momentum
    of the one-particle spectral function.}
    \label{fig:spectral}
\end{figure}

\begin{figure}
    \centering
    \includegraphics[scale=0.14]{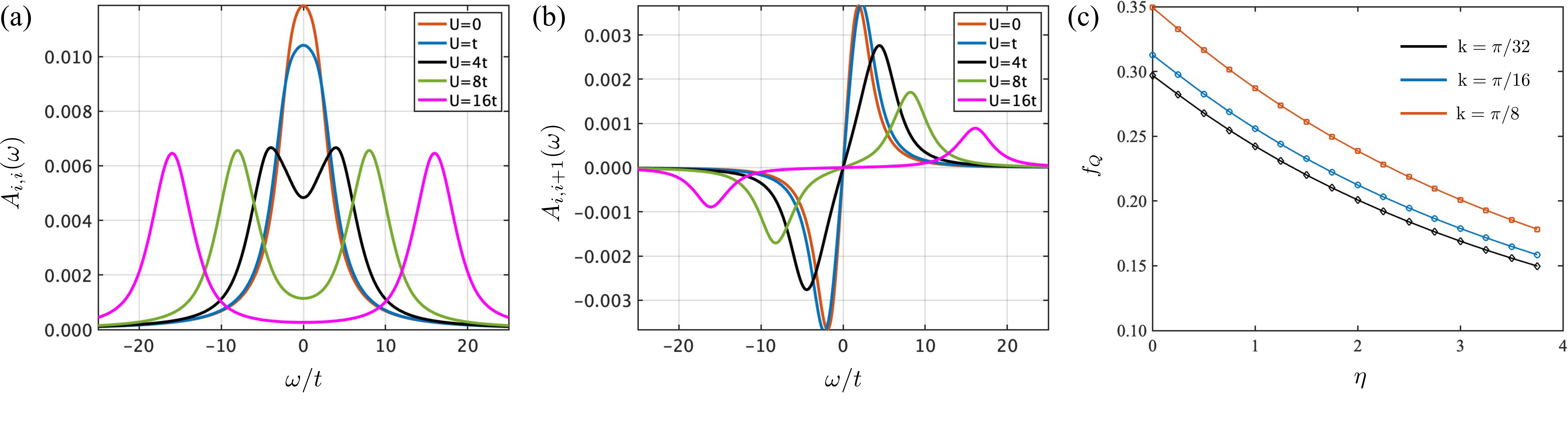}
    \caption{Position dependent one particle spectral functions are shown as a function of $\omega$ for the $t-U$ Hamiltonian for $U=4$ at $T=1$ in panels (a) and (b). The curves are derived by replacing $\delta$-functions with Lorentzians broadened by $\eta=2t$. (c) The QFI dependence on the broadening is shown for the same parameters as panels (a) and (b).}
    \label{fig:spectralposition}
\end{figure}

In the inset of Fig. 3 in the main text, we plot the fermionic one-particle spectral
function 
$A_{q}(\omega)$ for $T=1$. In Fig. \ref{fig:spectral} we show additional plots for the spectral functions at different temperatures $T=0$, $T=2$, and $T=4$. These spectral functions are computed by Fourier transforming the 
position-dependent functions $A_{i,j}(\omega)$, shown in Figs. \ref{fig:spectralposition}a, \ref{fig:spectralposition}b. The functions  $A_{i,j}(\omega)$ are broadened by 
$\eta=2t$, where we replace 
the delta function in \Eq{eq:spectral}
by a Lorentzian of width $\eta$, i.e., 
$\delta(x) \ \to\ \delta_\eta(x) \equiv 
\eta/[\pi(x^2+\eta^2)]$. In Fig. \ref{fig:spectralposition}c, we show the QFI dependence on the broadening parameter $\eta$ for three $k$ values of the witness operator and find that the broadening reduces the QFI. 
 
\section{Expressions for QFI of indefinite
electron number (IEN) wavefunctions\label{sec:2}}
Here we provide the general expression for the quantum Fisher information (QFI) in terms of the correlation matrix for wavefunctions that are not eigenstates of the total number operator.

Let the wavefunction of the system be $|\psi\rangle$.  We define the matrix $C$ and $P$ via
\begin{eqnarray}
    C_{ij} &=& \langle \psi |c^\dagger_i c_j|\psi\rangle;\cr\cr
    P_{ij} &=& \langle \psi |c^\dagger_i c^\dagger_j|\psi\rangle.
\end{eqnarray}
$C$ encodes number preserving fermionic correlations while $P$ encodes superconducting correlations.

In terms  of $C$ and $P$ the QFI for the witness operator in Eq.\,(2)
of the main text is
\begin{eqnarray}
  F_Q(|\psi\rangle,{\cal O}) =
  8\,{\rm Re} \sum_{i j}\Bigl[ a_i^\ast a_j^{\,}
  \bigl( \delta_{ij} C_{ii} - |C_{ij}|^2)
  + a_i^{\,} a_j^{\,}  |P_{ij}|^2 
  \Bigr] - 16\sum_i {\rm Re}(a_i)|\langle \psi|c^\dagger_i|\psi\rangle|^2
\text{ . }\notag \label{eq:D2O:2}
\end{eqnarray}
If the wavefunction $|\psi\rangle$ has an indefinite electron number but preserves electron parity, the last term in the above vanishes.

The equation (\ref{eq:D2O:2}) can be expressed in a compact form in terms of the extended correlation matrix $\mathcal{C}$ based on the spinor that combines all
creation and annihilation operators
$ \hat{\mathfrak{c}}^\dagger \equiv (\hat{c}_1^\dagger,
\hat{c}_2^\dagger, \ldots, \hat{c}_N^\dagger, 
\hat{c}_1^{\,}, \hat{c}_2^{\,}, \ldots, \hat{c}_N^{\,})$ 
into a $2N$ component vector. This defines 
$\mathcal{C}_{\rm i j} \equiv \langle \hat{\mathfrak{c}}_{\rm i}^\dagger
\hat{\mathfrak{c}}_{\rm j}^{\,}\rangle$ with ${\rm i,j} \in 1,\ldots,2N$
(note the roman script on ${\rm j}$ and ${\rm j}$ here),
or more compactly,
\begin{eqnarray}
   C \ \to\ \mathcal{C} \equiv
   \begin{bmatrix}
      C & P \\
      P^\dagger &1-C
   \end{bmatrix}
  \text{ .}\label{eq:CM:2}
\end{eqnarray}
This incorporates the pair correlation matrix $P$.
This is frequently used to describe fermionic Gaussian states
\cite{Eisert18,Surace22} or is obtained when computing
the plain correlation matrix from a purified mixed state.
For a physical correlation matrix, it must hold
$ \mathcal{C} (1- \mathcal{C}) \geq 0$
[completely analogous to $C(1-C) \geq 0$].
However, $\mathcal{C}$ satisfies additional constraints
as compared to the plain correlation matrix $C$ 
for a system of size $2N$ since, e.g., $P_{ii}=0$.

By also doubling $a \to \mathfrak{a} \equiv [a; -a^\ast]$
into a column vector of length $2N$,
one obtains the compact expression (for the case where electron number parity is preserved)
\begin{eqnarray}
  F_Q(|\psi\rangle, \mathcal{O}) =
  4\sum_{\rm i j}^{2N} \mathfrak{a}_{\rm i}^\ast \mathfrak{a}_{\rm j}^{\,}
  \bigl( \delta_{\rm i j} \mathcal{C}_{\rm i i} - |\mathcal{C}_{\rm i j}|^2)
\text{ .}\label{eq:D2O:3}
\end{eqnarray}
The lower diagonal block in \Eq{eq:CM:2}
could be taken as $C$ initially,
but it can be also written as $1-C$ since this does
not affect the variance in \eqref{eq:D2O:2}:
for $i\neq j$, $|C_{ij}|^2 = |-C_{ij}|^2$, and for the diagonal $i=j$,
with $n \equiv C_{ii}$ the average occupation of a site,
$n(1-n)$ is symmetric under $n \to 1-n$.

By writing the vector $\mathfrak{a} \to \mathcal{A} \equiv
{\rm diag}(\mathfrak{a})$, this allows one
to rewrite the witness operator in the form,
\begin{eqnarray}
   \hat{\mathcal{A}} \equiv
      \sum_{\rm ij}^{2N} \mathcal{A}_{\rm ij}
      \mathfrak{c}_{{\rm j}A}^\dagger \mathfrak{c}_{{\rm i}B}^{\,}
   = \sum_{i}^N \left(
    a_i \hat{c}_{iA}^\dagger \hat{c}_{iB}^{\,}
  - a_i^\ast \hat{c}_{iA}^{\,} \hat{c}_{iB}^\dagger 
  \right)
 = \hat{A} \,, \quad
\end{eqnarray}
which automatically includes the H.c. term.
With this \Eq{eq:D2O:3} can be rewritten in matrix notation
\begin{eqnarray}
  F_Q(|\psi\rangle,\mathcal{O})
  =8\, {\rm Re\ tr} \left[
    \mathcal{A} \mathcal{C} \bigl( \mathcal{A}^\dagger
    - \mathcal{A}^\ast \mathcal{C}^\dagger
  \bigr)\right]
\text{ .}\label{eq:D2O:4}
\end{eqnarray}%
Since \Eq{eq:D2O:3} represents a variance,
it must hold $F_Q \geq 0$. On general grounds then,
by splitting $i=j$ from $i\neq j$ in the sum \eqref{eq:D2O:3},
it must hold
\begin{eqnarray}
  F_Q(|\psi\rangle, \mathcal{O}) \leq
  8\sum_{\rm i}^{2N} |\mathfrak{a}_{\rm i}|^2
  \underbrace{\mathcal{C}_{\rm i i} (1 - \mathcal{C}_{\rm i i})}_{
    \leq 1/4}
  \leq 4 \sum_i^N |a_i|^2
\text{ .}
\end{eqnarray}
For $|a_i|^2=1$, this implies the upper bound $F_Q \leq 4N$.

\section{Discussion on the features of the maximal QFI for an  $N$ site block\label{sec:3}}
\subsection{Maximal QFI for systems with definite electron number (DEN)}
In this section, we discuss various features that are present in the maximal QFI curves shown in Fig. 1 of the main text. The Fig. 1 assumes fermionic systems at half-filling. The maximal QFI curves for $8$-partite and $4$-partite systems have wiggling features.  At first glance, this may seem like a numerical artifact. Upon close inspection, we observed that these wiggles are cosine functions. Furthermore, we observe that the maximal QFI reaches a strict maximum of $4N$ at $N-1$ points. An exact explanation of these features is beyond the scope of our analysis. However, some of these features can be intuitively understood from the expression for QFI (\ref{eq:D2O:2}) with DEN, i.e., with the $P$-matrix set to zero. Equation (\ref{eq:D2O:2}) can be rewritten in terms of cosine functions for a half-filled system of length N:
\begin{equation}
f_Q({\cal O}(k))=\frac{F_Q({\cal O}(k))}{N}=\frac{1}{2}-\frac{1}{N}\sum_{n} \cos{(n k)}\sum_{|i-j|=n} |C_{ij}|^2
    \label{eq:cos}
\end{equation}
Because for $N\geq 4$, the maximal QFI lies between $2N$ and $4N$, the second term in Eq. (\ref{eq:cos}) must be either positive or zero. 

\begin{figure}
    \centering
    \includegraphics[scale=0.25]{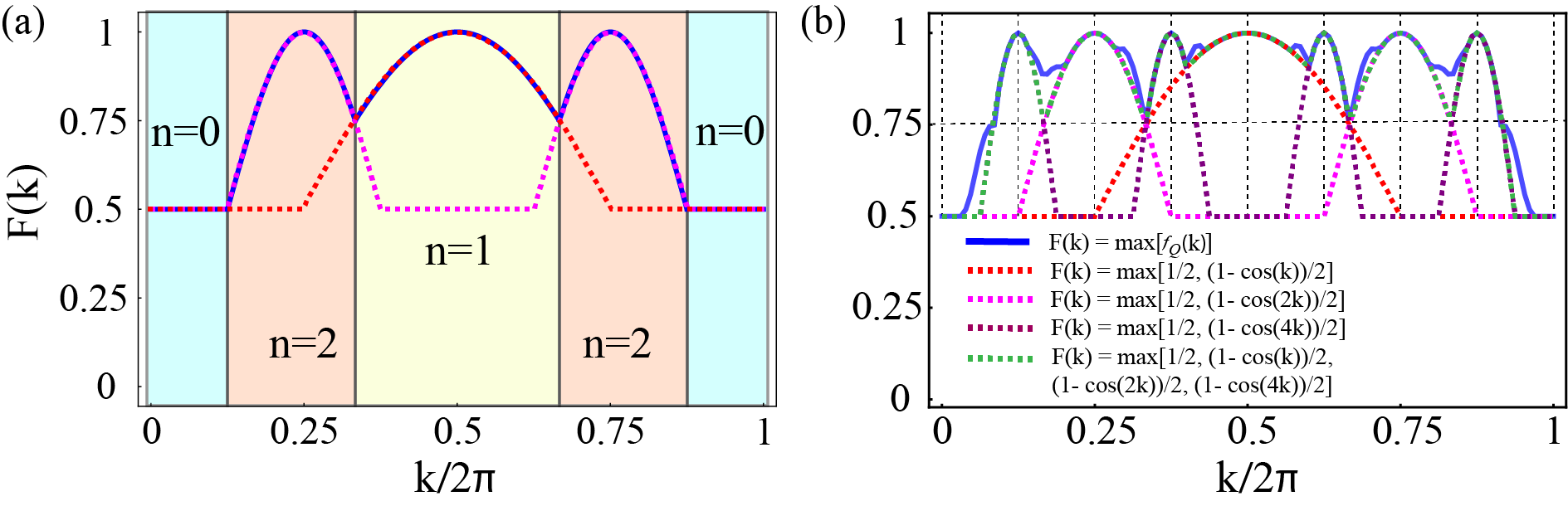}
    \caption{The maximal QFI for (a) $N=4$ and (b) $N=8$ site block is plotted as a function of $k$ together with functions where the QFI is computed from $C$ matrices where the off-diagonal elements $C_{ij}$ are confined to the condition $|i-j|=n$, where $n$ is an non-negative integer. (a) The yellow region corresponds to $n=1$, the orange region corresponds to $n=2$, and the green region represents a diagonal $C$ matrix $n=0$. (b) In addition to $n=1$, $2$ cases, the QFI for $N=8$ site includes the $n=4$ case which is not allowed in the $N=4$ case.}
    \label{fig:SM1}
\end{figure}

Let us consider the $N=4$ system at half-filling for specifity. We have observed that the $C$ matrix corresponding to a maximal QFI must have all the diagonal elements $1/2$. Then, for a particular $q$ value, the position of non-zero off-diagonal elements must ensure that the second part of the Eq. (\ref{eq:cos}) is positive. For small values of the wavevector $k$, the cosine functions in Eq. (\ref{eq:cos})  are close to $1$, and the second part of Eq. (\ref{eq:cos}) is negative for any non-zero off-diagonal elements. Therefore, up to finite values of k, the maximal QFI density is $2N$ as shown by the green region in Fig. \ref{fig:SM1}a. As we increase $k$, $\cos(nk)$ becomes negative for $n=3$ at $k>\pi/6$. However, such a $C$ matrix where only $C_{14}$ is non-zero is not physical for wavefunctions with DEN. Further increasing $k$, we observe that at $k>\pi/4$, $\cos(2q)$ becomes negative. This is shown by the orange region in Fig. \ref{fig:SM1}a which corresponds to a $C$ matrix where the non-zero off-diagonal elements are restricted to $|i-j|=n=2$. In the orange region starting from $k=2\pi/3$ $\cos(k)$ is negative and $\cos(k)<\cos(2k)$. At the intersection, $k=2\pi/3$, $\cos(k)=\cos(2k)=-1/2$. This sharp transition demonstrates that the non-zero off-diagonal elements shift from $n=2$ to $n=1$. This is further illustrated in the case of a $N=8$ site half-filled system, for which the maximal QFI density is shown with the QFI computed from a $C$ matrix with non-zero off-diagonal elements restricted to a single n, see Fig. \ref{fig:SM1}b.

\subsection{Maximal QFI for systems with IEN}
For wavefunctions with IEN, the $P$ matrix is non-zero in Eq. (\ref{eq:D2O:2}). The expression for QFI includes terms $\rm{Re}(a_i a_j)$ $\sim$ $\cos{k (i + j)}$, which depends on the sum of the positions. Consequently, while constant phase factors in $C$ occur
along off-diagonals $i-j = \rm n$,
constant phase factors in $P$ occur
along anti-diagonals, $i + j = \rm n$. The matrix elements $P_{ij}$, where  
\begin{eqnarray}
  i+j=N+1 \qquad (N \text{ even})
\label{eq:QFI:TIP:N+1}
\end{eqnarray}
is the only anti-diagonal that permits $N$
non-zero entries in $P_{ij}$ along $i=N+1-j \in [1,N]$
(assuming even $N$, the case $i=j$ does not occur).
For values $k=2m\pi/(N+1)$, where $m$ is a positive integer,  the maximum contribution from the $P$ matrix satisfies the condition $i+j=N+1$, so that the resulting cosine function $\cos{2m\pi}$ is $1$. Since, the QFI from the $C$ matrix alone does not give the strict maximum $4N$, the extended correlation matrix
$\mathcal{C}$ consisting solely of the diagonal and
antidiagonal (chosen maximal based on physicality)
permits the QFI to be maximized. 

\begin{figure}
\centering
\includegraphics[width=0.5\linewidth]{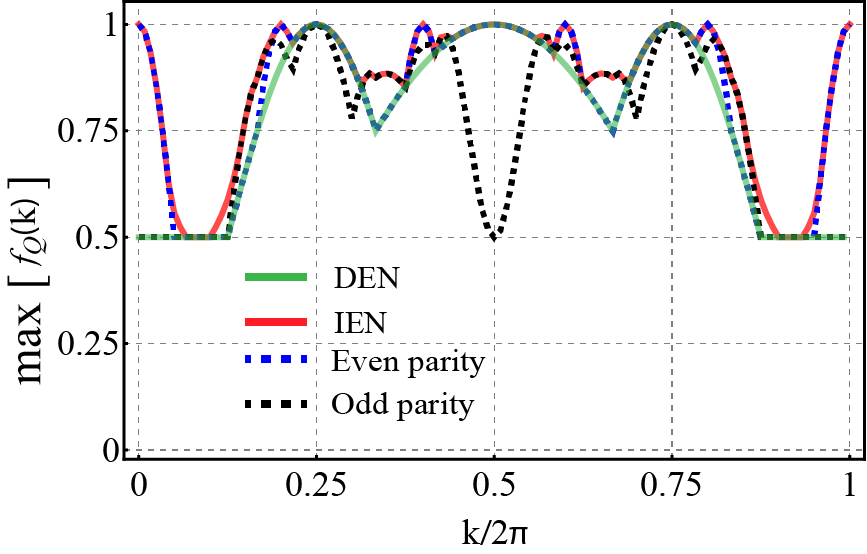}
\caption{Shown are the maximal QFIs for a 4-site fermionic spinless spinless system with and without charge conservation. The green curve corresponds to the case with a definite electron number (DEN) at half-filling and is the same as the curve for 4-partite entanglement in Fig.~1 in the main text. The red curve represents an indefinite electron number (IEN) while the two dashed curves represent the maximal QFI with an indefinite electron number but where the electron number parity is either even or odd.  
}\label{fig:symmetry}
\end{figure}

\section{Maximal QFI dependence on various symmetry constraints}
In the main text, we have only considered QFI maxima for wavefunctions that have DEN. In Fig. \ref{fig:symmetry} we explore relaxing this condition for an $N=4$ site system.  We consider the maximal QFIs for wavefunctions with an IEN and for wavefunctions that are eigenstates of electron number parity.
We see that IEN wavefunctions have a maximal QFI of $4N$ at the same $k$'s as DEN wavefunctions.  However, the IEN has a more complicated structure and additional values of $k$ where the maximal QFI is itself maximal, including $k=0$.  We can understand the increase in the maximal QFI at $k=0$ in going from DEN to IEN wavefunctions as a result of superconducting correlations being allowed as discussed in the previous section.  If we examine wavefunctions with IEN but where electron parity is a good quantum number, we see odd parity wavefunctions in general have smaller maximal QFIs that even parity wavefunctions apart from small regions on either side of $k/2\pi=1/8,7/8$.

Beyond questions of imposing electron number as a good quantum number upon the wavefunctions, we have examined whether the wavefunctions satisfying time reversal invariance (TRI), i.e., being real or complex, change the QFI maxima.  We have found it does not despite the parameter space of non-TRI wavefunctions being double that of TRI wavefunctions.

\section{Extending QFI Maximums to Larger Systems}

We have derived maximal QFI curves for different entanglement patterns for an $N=8$ site system.  However using the additivity of the QFI, the information in these maximal curves is applicable to larger systems.  Let us consider a system of size M.  Let us further consider a wavefunction $|\psi\rangle$ which has a multipartite entanglement pattern consisting of $K$ blocks where the i-th block involves a contiguous set of $M_i$ sites and where 
$\sum_{i=1}^K M_i = M$.  We correspondingly write the wavefunction as
\begin{equation}
    |\psi\rangle = \prod_{\otimes, i=1}^K|\psi_i\rangle.
\end{equation}
Additivity of the QFI
of our wavevector q-dependent witness operator that consists of a sum of local terms ${\cal O}(q)=\sum_ia_i(q){\cal O}_i$ then gives us
\begin{eqnarray}
    F_Q(|\psi\rangle,\sum_i a_i(q){\cal O}_i,M) &=& \sum_{j=1}^K F_Q(|\psi_j\rangle,\sum_{i\in M_{j}}a_i(q){\cal O}_i,M_j).
\end{eqnarray}
Now because the block of sites in $M_i$ is continguous, we can write $F_Q(|\psi_j\rangle,\sum_{i\in M_{j}}a_i{\cal O}_i,M_j)$ as
\begin{eqnarray}
    F_Q(|\psi_j\rangle,\sum_{i\in M_{j}}{\cal O}(q)_i,M_i) &=& \sum_{i,l\in M_j}(a_i(q)a^*_l(q)C_{il} + {\rm c.c.})\cr\cr
     &=& \sum_{i,l\in M_j}(a_{i-l}(q)C_{il} + {\rm c.c.})\cr\cr
     &=& \sum^{M_i}_{i,l=1}(a_{i-l}(q)C_{il} + {\rm c.c.}),
\end{eqnarray}
where here we have used the fact that the coefficients $a_i(q)a^*_l(q)$ depend on the site indices, $i$ and $l$, only through their difference $i-l$.  Thus if we maximize $F_Q(|\psi_j\rangle,\sum_{i\in M_{j}}{\cal O}_i,M_j)$ over all wavefunctions, we obtain the same maximal curve as a function of $q$ that we would obtain for an $M_i$ site system, i.e., the maximal curve does not care about the exact position of block of $M_i$ sites within the larger system with $M$ sites.  Thus
\begin{eqnarray}
    {\rm max} F_Q(|\psi\rangle,\sum_{i=1}^M a_i(q){\cal O}_i,M) = \sum_{j=1}^K{\rm max} F_Q(|\psi\rangle,\sum_{i=1}^{M_j} a_i(q){\cal O}_i,M_j).
\end{eqnarray}
Thus we can use the maximal QFI curves derived for smaller systems to construct the maximal curves for entanglement patterns in larger systems.  We note that this argument works only if electron number is a good quantum of the wavefunction.


\bibliography{bib,ref}